\documentclass[linenumbers,amsmath]{aastex631}
\usepackage{comment}
\usepackage{amsmath}
\usepackage{natbib}
\usepackage{xcolor}

\nolinenumbers

\begin{document}

\title{Median Statistics Estimate of the Distance to M87}

\author[0000-0002-4402-1343]{Nicholas Rackers}
\affiliation{Department of Physics, Washington University in St. Louis, Compton Hall, One Brookings Drive St. Louis, MO 63130}\thanks{E-mail: n.rackers@wustl.edu}
\author[0009-0001-7781-6265]{Sofia Splawska}

\affiliation{Department of Physics, Mellon College of Science, Carnegie Mellon University, 5000 Forbes Avenue
Pittsburgh, PA 15213}\thanks{E-mail: ssplawsk@andrew.cmu.edu}
\affiliation{Department of Physics, Kansas State University, 116 Cardwell Hall, Manhattan, KS 66506, USA}
\affiliation{Department of Physics, Case Western Reserve University, Rockefeller Building, 2076 Adelbert Road, Cleveland, OH 44106}
\author[0000-0002-7307-0726]{Bharat Ratra}
\affiliation{Department of Physics, Kansas State University, 116 Cardwell Hall, Manhattan, KS 66506, USA}\thanks{E-mail: ratra@phys.ksu.edu}


\begin{abstract}
\nolinenumbers
\cite{de_Grijs_2019} compiled 211 independent measurements of the distance to galaxy M87 in the Virgo cluster from 15 different tracers and reported $31.03 \pm 0.14$ mag as the arithmetic mean of a subset of this compilation as the best estimate of the distance. We compute three different central estimates --- the arithmetic mean, weighted mean, and the median --- and corresponding statistical uncertainty for the full data set as well as three sub-compilations. We find that for all three central estimates the error distributions show that the data sets are significantly non-Gaussian. As a result, we conclude that that the median is the most reliable of the three central estimates, as median statistics does not assume Gaussianity. We use median statistics to determine the systematic error on the distance by analyzing the scatter in the 15 tracer subgroup distances. From the 211 distance measurements, we recommend a summary M87 distance modulus of $31.08^{+0.05}_{-0.04}$ (statistical) $^{+0.04}_{-0.06}$ (systematic) mag, or combining the two errors in quadrature $31.08^{+0.06}_{-0.07}$ mag, rounded to $16.4 \pm 0.5$ Mpc, all at $68.27\%$ significance.
\end{abstract}



\section{Introduction} \label{sec:intro}

The extragalactic distance ladder is essential to astrophysics and cosmology and must constantly be refined. As the galaxy M87 lies near the center of the closest galaxy cluster to us, Virgo, it is an important rung on the distance ladder and allows us to extend the ladder to more distant clusters such as Coma and Fornax. \citet{de_Grijs_2019}, hereafter deGB, compiled a list of 211 distance measurements to M87 obtained by 15 different distance tracers. They report a mean value of $31.03 \pm 0.14$ mag after reducing their data set to 24 measurements from 3 tracers which they believe to be well-calibrated and independent.

Using the mean as a central estimate, as deGB did, implicitly assumes Gaussianly distributed data, or at least a distribution with deviations from an underlying symmetry and outliers only at the noise level. However, we determine that the full compilation, as well as the two sub-compilations studied by deGB are significantly non-Gaussian.\footnote{\cite{Ramakrishnan:2023vlq} have more thoroughly examined the Gaussianity of the full deGB compilation, as well as that of many of the 15 different tracer compilations, and also find that the full deGB compilation is non-Gaussian.} Non-Gaussian data compilations are not uncommon in astronomy: well known examples including Hubble constant measurements \citep{Gott_III_2001, Chenetal2003, Chen_2011, Calabreseetal2012, BethapudiDesai2017, zhang2018most}, cosmological mass density estimates \citep{ChenRatra2003}, distances to the SMC and LMC \citep{Crandall_2015, de2014clustering}, and measurements of the Solar radius and Galactic rotational velocity \citep{camarillo2018median, Camarillo_2018, RajanDesai2018, BobylevBajkova2021, de2016clustering, de2017clustering}. For further examples see \citet{Crandalletal2015}, \citet{Bailey2017}, \citet{Zhang2017}, \citet{RajanDesai2020}, and \citet{Zhangetal2022}.

Following \citet{crandall2014median}, \citet{Penton_2018}, and \citet{yu2020gaussian}, we analyze the deGB data sets using median statistics \citep{Gott_III_2001}, which is free from assumptions about the distribution underlying the data set and its errors. Because median statistics does not take into account error bars on individual measurements, it is generally less constraining. Nevertheless, we believe the median to provide a more accurate estimate of the distance to M87 than methods which rely on the unsatisfied assumption of Gaussianity. In addition to using median statistics to estimate a more reliable statistical uncertainty in the M87 distance measurement, we also utilize it to estimate the systematic uncertainty in this measurement, based on the scatter in the M87 distance estimated using each of the 15 different tracer sub-groups in the deGB compilation.\footnote{\citet{Ramakrishnan:2023vlq} have also performed a median statistics analysis of the deGB compilation, not based on the \citet{Gott_III_2001} technique but on one that assumes only mildly non-Gaussian data and so get a different statistical error on the median. They do not estimate the systematic error on the median.}    

The results of this median statistics analysis on the entire data set of 211 distance measurements provided by deGB yields an M87 distance of $31.08^{+0.04}_{-0.04}$ (statistical) $^{+0.04}_{-0.06}$ (systematic) mag at $68.27\%$ significance. Combining the two errors in quadrature we get $31.08^{+0.06}_{-0.07}$ mag or $16.4 \pm 0.5$ Mpc.

In Section \ref{sec:data} we introduce the different data compilations studied. In Section \ref{sec:analysis} we summarize median statistics and outline the Gaussianity test we use. In Section \ref{sec:results} we study the Gaussianity of the deGB compilations and argue that our median statistics result is a better representation of the true distance to M87 than a more conventional mean analysis. We conclude in Section \ref{sec:conclusion}.

\section{Data} \label{sec:data}

deGB compiled 211\footnote{deGB uses 213 measurements, but their tracer organized database only lists 211. The missing 2 points do not statistically affect the results.} M87 distance measurements from the NASA/Astrophysics Data System (ADS) that they found to be statistically independent. They included measurements both to M87 and to the geometric center of the Virgo cluster since these values were found to be statistically indistinguishable. These were grouped into 15 tracers.\footnote{Refer to Table \ref{tab:m87_systematic} for a full list.} deGB selected 5 of these tracers: Cepheids, planetary nebulae luminosity function (PNLF), surface brightness fluctuations (SBF), tip of the red giant branch (TRGB) magnitude, and novae, which they found to be internally consistent and provide tight averages as opposed to the other 10 tracers. These 5 tracers correspond to a total of 44 measurements. They adjusted the measurements from these 5 tracers to agree with their best estimate of the distance modulus to the LMC of $18.49 \pm 0.09$ mag \citep{de2014clustering}.\footnote{The unadjusted measurements database sorted by tracer type can be found at \hyperlink{https://astro-expat.info/Data/m87distbytracer.html}{https://astro-expat.info/Data/m87distbytracer.html}.} This value is in good agreement with the median statistics estimate of $18.49 \pm 0.13$ mag found by \cite{Crandall_2015}. deGB's final recommended value is the arithmetic mean of a set of 24\footnote{deGB say they use 28 data points, so 27 after they discard the \citet{tammann2000distance} point (as discussed next); however, they list only 25, including the \citet{tammann2000distance} point, in their Table 1, possibly due to a confusion between their unadjusted data-points and adjusted data-points, of which there are 3 fewer. deGB discard one particular outlier of the Cepheid tracer from \citet{tammann2000distance}, who reported this result under the assumption of the long distance scale. Currently, the short distance scale is the accepted framework \citep{freedman2001final}.} points from three tracers: Cepheids, SBF, and TRGB. The PNLF and novae tracers were removed due to foreground- and background-biased outliers, respectively. The result of deGB's mean analysis of the 24 points from these 3 tracers is a distance modulus of $31.03 \pm 0.14$ mag at 1$\sigma$ sample standard deviation; we find an identical result. 

In this paper we analyze the following four subsets of the deGB compilation and provide the best estimate for each subset with 1$\sigma$ uncertainty as explained in Section \ref{sec:central_estimate}. All 15 refers to deGB's full list of 211 unadjusted data points from 15 tracers. All 15 without averages is the same but now excluding the averages tracer.\footnote{So that we may determine what effects removing this tracer has on the results.} Best 5 refers to  the adjusted 44 measurements from 5 tracers from Table 1 of deGB that they determined to be internally consistent. Best 3 refers to the adjusted 24 data points from 3 tracers \citep[excluding the][point]{tammann2000distance} from Table 1 of deGB that they used to compute their favored summary measurement value. 

\section{Analysis} \label{sec:analysis}

Conventional methods such as mean and $\chi^2$ analyses assume \citep{Gott_III_2001}:
 
\begin{enumerate}
    \item Individual data points are statistically independent.
    \item There are no systematic effects.
    \item The errors are Gaussianly distributed.
    \item One knows the standard deviation of the errors. 
\end{enumerate}

Median statistics was developed by \cite{Gott_III_2001} as a powerful alternative to mean and $\chi^2$ analyses. The essential idea is that the true value of the quantity being measured is the median of a set of repeated error-affected measurements as the number of measurements tends toward infinity. This follows from the assumptions that the data set contains only independent measurements and that it does not have any overall systematic error.\footnote{There can be systematic error in different subsets of the data set, so long as the same systematic error does not affect a significant portion of the full data set.} The individual measurement errors have no effect on the computation of the median of a data set. Thus assumptions 3 and 4 are not necessary for the application of median statistics. This is advantageous for analyzing non-Gaussian data sets, as for such cases any mean analysis is suspect due to the failure to satisfy assumption 3. Since the individual measurement errors are not taken into account (assumption 4 is dropped), a median statistics analysis will generally provide a less constraining central estimate than mean statistics. We argue that despite this, median statistics provides the most reliable central estimate in the case of a non-Gaussian data compilation.

We study the Gaussianity of the compilations described in Section~\ref{sec:data}. We do this by creating error distributions of the data from various central estimates and comparing them to the Gaussian probability distribution. Based on the results of this analysis, we argue that the median is the most accurate and reliable central estimate of the deGB compilations.

\subsection{Computing the Central Estimate}
\label{sec:central_estimate}

To study the Gaussianity of a data set, we construct an error distribution using a central estimate. We compute the median, weighted mean, and arithmetic mean central estimates and create an error distribution for each. 

The true median of a data set is defined as the median as the number of measurements, $N$, goes to infinity. \cite{Gott_III_2001} showed that the probability that the true median lies between any two adjacent measurements $M_i$ and $M_{i+1}$ is given by the appropriately normalized binomial distribution:
\begin{equation}
    P = \frac{2^{-N}N!}{i!(N-i)!},
    \label{eq:med_prob}
\end{equation}
where $M_0 	\equiv -\infty$ and $M_{N+1} 	\equiv +\infty$. To compute the uncertainty in our estimate of the median: (i) for every index, we find the probability that the true median is between it and the subsequent index; (ii) we split the data into two halves above and below the median of our distribution; (iii) we compute the area of each half; (iv) we iterate from the median until $68.27\%$ of the area of a half is exceeded; and, (v) the data at the index which yields an area closest to the $68.27\%$ is then recorded as $M_{i+}$ or $M_{i-}$.

These indices are then used in the ordered data to construct the uncertainty as described:
\begin{align*}
    \sigma_+ &= M_{i+}-M_{\rm med}, \\            
        \sigma_- &= M_{\rm med} - M_{i-},
\end{align*}
where $M_{\rm med}$ is the median.

The weighted-mean central estimate, while it has the added assumption of Gaussianity, has the benefit of using the reported uncertainties and so is a more constraining estimate than the median. Given a data set $M_i \pm \sigma_i$ we compute the weighted-mean central estimate as \citep{Podariuetal2001}
\begin{equation}
    M_{\rm wm} = \frac{\sum_{i=1}^{N}M_i / \sigma_i^2}{\sum_{i=1}^{N}1/\sigma_i^2},
    \label{eq:weighted_mean}
\end{equation}
with standard deviation
\begin{equation}
    \sigma_{\rm wm} = \frac{1}{\sqrt{\sum_{i=1}^{N}1/\sigma_i^2}}.
    \label{eq:wm_std}
\end{equation}

We also employ the arithmetic mean central estimate 
\begin{equation}
    M_{\rm m} = \frac{1}{N}{\sum_{i=1}^{N}M_i},
\end{equation}
with standard error of the mean
\begin{equation}
    \sigma_{\rm m} = \sqrt{\frac{1}{N^2}{\sum_{i=1}^{N}(M_i - M_m)^2}}.
\end{equation}

These central estimates are used to construct error distributions which we will compare with a Gaussian in order to evaluate the Gaussianity of the data set.


\subsection{Error Distributions}

We create error distributions of our data in order to study their Gaussianity. The error distribution $N_{\sigma_i}$ is a measure of how many standard deviations any individual measurement deviates from the central estimate. 

Given measurements $M_i \pm \sigma_i$ and a corresponding central estimate $M_{\rm CE} \pm \sigma_{\rm CE}$, the error distribution is
\begin{equation}
    N_{\sigma_i} = \frac{M_i - M_{\rm CE}}{\sqrt{\sigma_i^2 + \sigma_{\rm CE}^2}}.
    \label{eq:err_dist}
\end{equation}
This expression assumes that the central estimate is not correlated with the data set. This assumption is not satisfied here, as our central estimates are computed directly from the data compilations. Finding such a formula for an error distribution of a correlated median is beyond the scope of this work.

The weighted-mean case has been solved in the case of correlation between the weighted-mean and the data set.\footnote{A derivation is shown in \cite{Camarillo_2018}.} The error distribution using a weighted-mean central estimate that is correlated with the data set is
\begin{equation}
    N_{\sigma_i} = \frac{M_i - M_{\rm wm}}{\sqrt{\sigma_i^2 - \sigma_{\rm wm}^2}}.
    \label{eq:wm_err_dist}
\end{equation}

We often have asymmetric error bars on the median, in which case we slightly alter Equation (\ref{eq:err_dist}). We use the upper error of the median when $M_i > M_{\rm med}$ and the lower error when $M_i < M_{\rm med}$
\begin{equation}
N^{\rm med}_{\sigma_i} = 
    \begin{cases}
    \begin{aligned}
    \frac{M_i - M_{\rm med}}{\sqrt{\sigma_i^2 + \sigma_{\rm med}^2}}, && M_i \geq M_{\rm med}\\
    \frac{M_i - M_{\rm med}}{\sqrt{\sigma_i^2 - \sigma_{\rm med}^2}}, && M_i < M_{\rm med}. 
    \end{aligned}
    \end{cases}
\end{equation}

After symmetrizing the error distribution about $0$,\footnote{For example: [1, 2, 3] $\rightarrow$ [$-3$, $-2$, $-1$, 1, 2, 3].}
    \label{eq:sym_err_dist} we use the Kolmogorov-Smirnov (KS) test to study the Gaussianity of the data compilations.


\subsection{The Kolmogorov-Smirnov Test and Testing Gaussianity}

The KS test is a statistical test that compares empirical error distributions with continuous probability density functions (PDFs).\footnote{We also used the Anderson Darling (AD) test for this purpose. The AD test results agree with our KS test findings and so are not recorded here.} The first step is calculating the $D$-statistic, which is the largest difference between the empirical cumulative distribution function and that of the relevant PDF. 

The $D$-statistic is then used to compute $z$ as given in \cite{press2007numerical}
\begin{equation}
    z = \left(\sqrt{N} + 0.12 \frac{0.11}{\sqrt{N}}\right)D,
    \label{eq:z_val}
\end{equation}
which is used to compute the $p$-value
\begin{equation}
    p = 2 \sum_{i=1}^{\infty}(-1)^{i-1}e^{-2i^2z^2}.
    \label{eq:p_val}
\end{equation}
Explicitly, the $p$-value is the probability that the $D$-statistic could be smaller than measured if a similar data set is used. The $p$-value represents the probability that we can reject the null hypothesis which states that these data do not come from the PDF of interest. Conventionally, if $p \geq 0.95$, we can reject the null hypothesis that our data do not come from the relevant PDF. Therefore, if $p \leq 0.95$, we conclude that these data are not consistent with having been drawn from a Gaussian distribution.

We also introduce a scale factor $S$ such that
\begin{equation}
    N_{\sigma_i}^{\rm scaled} = \frac{N_{\sigma_i}}{S}.
    \label{eq:scale}
\end{equation}
$S > 1$ corresponds to decreasing the errors, or narrowing the distribution, and $S < 1$ corresponds to increasing the errors, or widening the distribution. We then run the KS test, varying $S$ from 0 to 10 to find $S^*$, the value of $S$ which optimizes the $p$-value. This allows us to compare the error distribution to a Gaussian. If $S^* > 1$, that is, if the optimal scale factor is such that the distribution must be narrowed to fit a Gaussian, the distribution is broader than a Gaussian and we conclude that the errors may have been overestimated. Similarly, if $S^* < 1$, the distribution is more narrow than a Gaussian and we conclude that the errors may have been underestimated.


\subsection{Estimating Systematic Error}

All uncertainties computed thus far have corresponded to statistical errors. We perform an analysis of the systematic error present in the All 15 data set, and in the All 15 without Averages data subset, using the procedure outlined in \cite{Chen_2011}. Within each tracer subgroup there is statistical uncertainty resulting in a spread of measurements, and between the tracers there could be systematic error resulting from the different techniques and calibrations.

We construct a new data set consisting of the median of each tracer subgroup. We perform a median statistics analysis on this new data set to find the median of medians and its associated uncertainty. If we assume that these medians differ only systematically from each other, this uncertainty corresponds to the systematic uncertainty of the entire group of tracers. 


\section{Results} \label{sec:results}

We perform a median statistics analysis on deGB's compilation of 211 distance measurements. We calculate three central estimates and study the Gaussianity of the All 15 data set (comprised of 211 measurements from 15 tracers), the Best 5 data set (comprised of 44 adjusted measurements from 5 tracers), and the Best 3 data set (comprised of 24 adjusted measurements from 3 tracers). These data sets are outlined in Section \ref{sec:data}. The central estimates and results of the KS test for these data are shown in Table \ref{tab:central_estimates}.

\begin{table}[ht]
    \caption{Central estimates and KS test results for the four data compilations.}
    \centering
    \begin{tabular}{l|c|c|c|cc}
    \hline\hline
         Data Set & & Central Estimate (mag) & $p$\footnote{The $p$-values in this column are for the unscaled error distribution, or $S = 1$.} & $S^*$ & $p^*$\footnote{The 1.0 values in the $p^*$ column all lie within the range of (0.995, 0.999).} \\\hline
         & Median         &  $31.08^{+0.04}_{-0.04}$   & $\ll 0.1$  &2.2 & 0.83      \\
         All 15 & Weighted Mean\footnote{There were some data without errors. These were set to the mean of the uncertainties for that tracer in order to perform the weighted mean analysis.}  &  31.07       $\pm 0.01$&$\ll 0.1$ &2.4 & 0.63      \\
         & Arithmetic Mean&  30.97        $\pm 0.07$& $\ll 0.1$&2.1 &0.51      \\
         \hline\hline
        All 15& Median       & $31.08^{+0.05}_{-0.04}$     & $\ll 0.1$ &2.1&0.85    \\
        w/o Averages&Weighted Mean & 31.06 $\pm0.01$&$\ll 0.1$&2.1&0.68 \\
        &Arithmetic Mean& 30.95$\pm0.08$ &$\ll 0.1$ & 1.9 & 0.49 \\
         \hline\hline
         & Median         &$31.02^{+0.04}_{-0.05} $        & 0.20      &1.4 &1.0       \\
         
         Best 5\footnote{Including the \citet{tammann2000distance} measurement.} & Weighted Mean  &31.03        $\pm 0.01$     & 0.039  &1.8  & 1.0     \\
         & Arithmetic Mean&31.04         $\pm 0.03$     & 0.057        &1.6 & 1.0      \\
         \hline\hline
         & Median         &$31.025^{+0.04}_{-0.005}  $    &    0.26      &1.7 &1.0       \\
         Best 3\footnote{Excluding the \citet{tammann2000distance} measurement.} & Weighted Mean  &31.05        $\pm 0.01$     & 0.109  &1.9  & 1.0     \\
         & Arithmetic Mean&31.03         $\pm 0.03 $    & 0.30               &1.8 &1.0     \\
         \hline\hline
    \end{tabular}
    \label{tab:central_estimates}

\end{table}

We find all four data sets to be inconsistent with Gaussianity. Therefore, we argue that the median provides the best central estimate of each of these compilations. As the optimal scale factor $S^* > 1$ for all of these data sets, the error distributions are all wider than a Gaussian. This could be due to an overestimation of some of the errors.

We report a median of $31.08^{+0.04}_{-0.04}$ mag for the All 15 data set, $31.08^{+0.05}_{-0.04}$ for the All 15 without averages data set, $31.02^{+0.04}_{-0.05}$ mag for the Best 5 data set, and $31.03^{+0.04}_{-0.01}$ mag for the Best 3 data set, where all errors are at a $68.27\%$ confidence level.

In the case of the All 15 and All 15 without averages data sets, there are enough tracers to estimate the systematic error in the median of the entire data as outlined in Section \ref{sec:analysis}. The results of this analysis are shown in Table \ref{tab:m87_systematic}. The median of the All 15 data set is $31.08^{+0.04}_{-0.04}$ (statistical) $^{+0.04}_{-0.06}$ (systematic) mag and the median of the All 15 without averages is $31.08^{+0.05}_{-0.04}$ (statistical) $^{+0.04}_{-0.06}$ (systematic) mag, at $68.27\%$ significance. The systematic error due to the tracer differences is comparable to the statistical uncertainty on the median. Combining the two errors in quadrature we get $31.08^{+0.06}_{-0.07}$ mag or $16.4 \pm 0.5$ Mpc for both the All 15 and the All 15 without averages data sets.

\begin{table}[ht]
\centering
\caption{All 15 and All 15 without Averages Systematic Uncertainty Analyses.\footnote{All Data are the 211 data points including the \citet{tammann2000distance} measurement. The next 15 lines are the individual tracer types. Only for tracer types with more than 10 measurements do we show their uncertainty in the last column. The Subgroup Medians row shows the results from a median statistics analysis of the previous 15 medians and its uncertainty is the reported systematic uncertainty.}}
\begin{tabular}{lccccc}
\hline\hline
 Tracer Type      &  $N$  & Median (Shift)\footnote{The shift is the difference between the tracer median and the All Data median in the first row.}& w/o Ave (Shift)   & 1$\sigma$ Error (Width)\footnote{Error on the median of the previous column.}  & w/o Ave 1$\sigma$ Error (width)  \\
\hline
 All Data         & 211 & 31.08 &             & $-0.04$, +0.04 (0.08)    & $-0.05$, +0.05 (0.09)   \\
 TFR............. & 36  & 31.24 (0.16) &(0.16)   & $-0.16$, +0.16 (0.32)   & $-0.16$, +0.16 (0.32)  \\
 GCLF............ & 32  & 31.11 (0.03)&(0.04)     & $-0.14$, +0.04 (0.18)   & $-0.14$, +0.04 (0.18)  \\
 Averages........ & 21  & 31.08 (0.00) & ---      & $-0.02$, +0.15 (0.17)    & --- \\
 SBF............. & 18  & 31.12 (0.04)&(0.05)     & $-0.09$, +0.03 (0.12)   & $-0.09$, +0.03 (0.12)  \\
 SNe............. & 18  & 31.65 (0.57) & (0.57)   & $-0.05$, +0.05 (0.10)   & $-0.05$, +0.05 (0.10)  \\
 Other Methods & 15  & 30.90 ($-0.18$)  & ($-0.18$)   & $-0.10$, +0.25 (0.35)   &  $-0.10$, +0.25 (0.35)  \\
 PNLF............ & 12  & 30.87 ($-0.22$) &($-0.21$)  & $-0.02$, +0.03 (0.05)  &  $-0.02$, +0.03 (0.05)  \\
 Faber-Jackson    & 11  & 31.14 (0.06)  & (0.07)    & $-0.01$, +0.34 (0.35)    &  $-0.01$, +0.34 (0.35)   \\
 Color-magnitude & 11  & 30.84 ($-0.24$) & ($-0.24$)    & $-0.08$, +0.06 (0.14)  &  $-0.08$, +0.06 (0.14)   \\
 Novae........... &  8  & 31.40 (0.32)  & (0.33)    & ...                   &  ... \\
 Hubble law...... &  8  & 27.30 ($-3.78$) & ($-3.78$)   & ...               &   ...    \\
 Cepheids........ &  7  & 31.02 ($-0.06$) & ($-0.06$)   & ...               &  ...     \\
 HII............. &  6  & 31.20 ($0.12$) & (0.13)     & ...                &   ...   \\
 Group Member &  5  & 30.50 ($-0.58$)     &($-0.58$)   & ...                &   ...   \\
 TRGB............ &  3  & 31.05 ($-0.03$) & ($-0.03$)   & ...               &   ...    \\
 \hline
Subgroup Medians &   15   & 31.08 & 31.08          & $-0.06$, +0.04 (0.1)
& $-0.06$, +0.04 (0.1)  \\
\hline\hline
\end{tabular}
\label{tab:m87_systematic}
\end{table}

 The All 15 data set has the most measurements, enough to allow an estimate of systematic uncertainty, and also to make this data-set the best defended against the effects of small number statistics. Therefore, we choose to use the results of the analysis of this data-set as our final reported value.


\section{Conclusion} \label{sec:conclusion}

After analyzing the data sets compiled by deGB, we recommend an M87 median statistics distance modulus from the All 15 data set without averages of value $31.08^{+0.05}_{-0.04}$ (statistical) $^{+0.04}_{-0.06}$ (systematic) mag at $68.27\%$ significance. Combining the two errors in quadrature we have $31.08^{+0.06}_{-0.07}$ mag or $16.4 \pm 0.5$ Mpc. This estimate is consistent with deGB's result of $31.03 \pm 0.14$ mag based on the Best 3 data set. We argue that our reported value is more reliable than deGB's mean statistics analysis since these data are not Gaussianly distributed. Using the larger data set allowed us to estimate systematic uncertainty and include that in our reported value.

\section{Acknowledgments}
We acknowledge helpful discussions with Jacob Peyton, Aman Singal, Shantanu Desai, and Gunasekar Ramakrishnan. This project was supported by funding from Kansas State University’s REU program funded by NSF grant No.\ 1757778.

\bibliography{Median_Analysis}{}
\bibliographystyle{aasjournal}

\end{document}